\documentstyle[epsfig]{aipproc}

\begin{document}
\title{Is Positron Escape Seen In the\\ Late-time Light Curves of\\
 Type Ia Supernovae?}

\author{Peter A. Milne, Lih-Sin The, Mark D. Leising}   
\address{Department of Physics and Astronomy, Clemson University,
Clemson SC 29634-1911}

\maketitle

\begin{abstract}
At times later than 200 days, Type Ia SN light curves are
dominated by the kinetic energy deposition from positrons
created in the decay of $^{56}$Co.
In this paper, the transport of positrons after emission are simulated,
for deflagration and delayed detonation models, assuming
various configurations of the magnetic field.
We find the light curve due to positron kinetic energy has a different
shape
for a radially combed magnetic field
than it does for a tangled field that traps positrons.
The radial field light curves fit the observed light curves better than
do the trapping field light curves for the four SNe used in this study.
The radial field permits a much larger fraction of the positrons to escape,
perhaps enough to explain a large percentage
of the 511 keV annihilation radiation
from the Galactic plane observed by OSSE.
\end{abstract}

\section*{Introduction}

Type Ia supernovae possess two qualities that make them 
attractive as a potential
source of positrons; the large amount of positron-producing
$^{56}$Ni synthesized in the explosion, and the occurrence of Type Ia SNe
in bulge populations. Their spectra and light curves show evidence of
the $^{56}$Ni $\rightarrow$ $^{56}$Co $\rightarrow$ $^{56}$Fe decays. In
19\% of the $^{56}$Co $\rightarrow$ $^{56}$Fe decays, a positron is
produced. The large number of positrons produced in a single SN 
combined with realistic Type Ia SN rates show that they can easily
account for the 511 keV annihilation flux observed by SMM and other
detectors if enough can escape from the SN into the ISM. All models of
the galactic distribution of 511 keV annihilation emission that fit the
OSSE observations contain a bulge component. Supernova searches in other
spiral galaxies confirm that Ia SNe occur in the bulge populations
\cite{branch96}.

Heterogeneity does exist within the Ia paradigm. Different 
models are created to
fit the light curves and spectra of individual SNe. The mass of
$^{56}$Ni produced in models varies from 0.1 
 -0.9 M$_{\odot}$
\cite{hoeflich96}. The amount and location of $^{56}$Ni in the ejecta
has an obvious effect upon the fraction of positrons that escape. The
magnetic field geometry is a less obvious, but even more important
determinant of positron escape. If the field is tangled such that 
positrons are eternally trapped within the ejecta, the fraction of
positrons that escape is zero. A small fraction may survive
non-thermally in the expanding ejecta ($\leq$10$^{-3}$ survival fraction).
If the homologous expansion stretches the
 field lines to the extent that they are
essentially radial, then many more positrons may escape (5\%
-15\%). This subject was explained in detail by Chan and Lingenfelter
(1993)\cite{chan93}.
 While extreme SN Ia models vary in nickel mass by a factor
of less than 10,
the ratio between positron survival in the radial field geometry
and positron survival in the knotted field geometry can vary by 
a factor of 200.
Thus, the field geometry is the dominant
determinant of positron survival. The problem is to determine which
field geometry approximates the SN ejecta the best.

\section*{Light Curves}

Six times as much energy goes into the $\gamma$-rays 
($\simeq$3.6 MeV per decay)
 as goes into the
kinetic energy of the positron ($\simeq$0.632 MeV)
\cite{chan93}
For positron KE deposition to play a significant role, the
gamma-ray opacity must drop to below one-sixth the effective opacity of
the positron KE deposition. The gamma-ray transport was
treated with a Monte Carlo simulation
which estimated the fraction ($f_{dep,\gamma}$)
 of the decay energy that is deposited via
Compton scattering and photoelectric absorption processes.
$f_{dep,\gamma}$ is published for a number of SN models, including W7, in
The, Bridgeman, Clayton (1994)\cite{the94}. 
The radial positron KE deposition
was calculated by emitting positrons isotropically relative to the field
lines
 and slowing them via ionization
and excitation, inverse Compton scattering, synchrotron emission, and
bremsstrahlung emission. Ionization and excitation
dominates at all but the highest energies. 
The trapped 
positron KE deposition was calculated by emitting the
positrons in the center of their respective zones and confining them to
evolve co-located with the ejecta. 

\begin{figure}[t!] 
\centerline{\epsfig{file=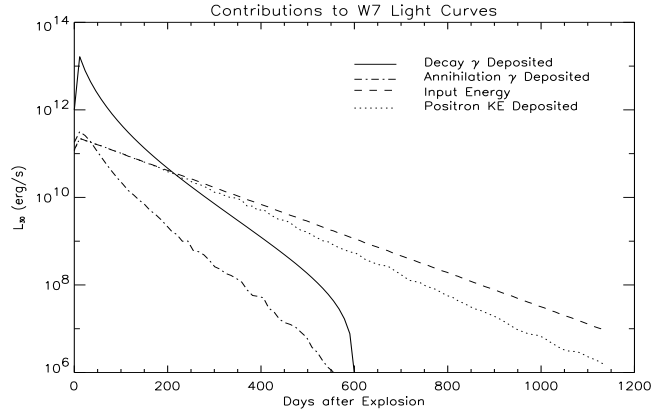,height=2.2in,width=3.5in}}
\vspace{10pt}
\caption{$\gamma$ and e$^{+}$ KE deposition in units of 10$^{30}$ erg/s.
The ``input energy" is the KE input to the ejecta by $\beta^{+}$ decay.
 The ``In Situ" approximation is the assumption that all of the KE is
deposited on-site, instantaneously. The  energy deposited by 
$\gamma$-rays produced by
positron annihilation
in the ejecta is not significant.}
\label{fig1}
\end{figure}

Figure 1 shows the various contributions to the bolometric light curve of
the SN Ia model W7\cite{nomoto84}.
The opacity to $\gamma$-rays drops
quickly, so by 200$^{d}$ only a small fraction of
the $\gamma$-rays interact with the
ejecta, and the KE deposited by the positrons as they 
slow down dominates the light
curve. 
Escaping positrons take some of
their KE with them creating a deficit in the light curve versus the ``In
Situ" approximation. Trapped, non-thermal positrons will store energy
temporarily, giving it back at later times.
The radial curve falls off more sharply than does 
the In Situ curve; the trapped
curve crosses over above the In Situ curve and remains slightly above
that curve. 

\section*{Observations}

A few SNe have occurred close enough that photometry was taken at
350+ days. A model is judged
to satisfactorily
 represent the observed supernova if it reproduces the early-time
spectrum and the multi-band light curve. The test of the field geometry is achieved 
by using models which fit at early time and determining whether
the predictions of either of the two field geometries fit the late-time
B band
light curve. The B band is assumed to be proportional to the bolometric
light curve during this phase. The bases of this assumption are two-fold.
First, 
the observed light curves do not show color evolution in this phase
\cite{turatto90}.
Second, once the
 nebular phase begins, the continuum falls off
 and the B band is dominated by florescence lines\cite{colgate97}.
 The lines should scale with the
KE deposition until very late times ($\sim$ 1000$^{d}$) at which time
freeze out becomes an issue\cite{schmidt94}. Thus, the phase to observe positron driven
light curves is between 400 and 1000 days.

The light curves are fit on a supernova-by-supernova basis.
Four SNe were analyzed in this preliminary study. The first two, SN
1972E\cite{kirshner75}
 and SN 1992A\cite{suntzeff97} were considered 
"typical" SNe and are thus fit with
the deflagration model 
W7 to demonstrate the technique. The second two, SN 1990N\cite{lira95}
 and SN 1991T\cite{schmidt94}
were peculiar and are fit by late detonation
models created specifically to fit their early spectra and light curves,
W7DN and W7DT\cite{yamaoka92}. The lines are scaled to fit the data
at 200 days.  The fits are shown in figure 2.

\begin{figure}[t!] 
\makebox[2.6in]{\epsfig{file=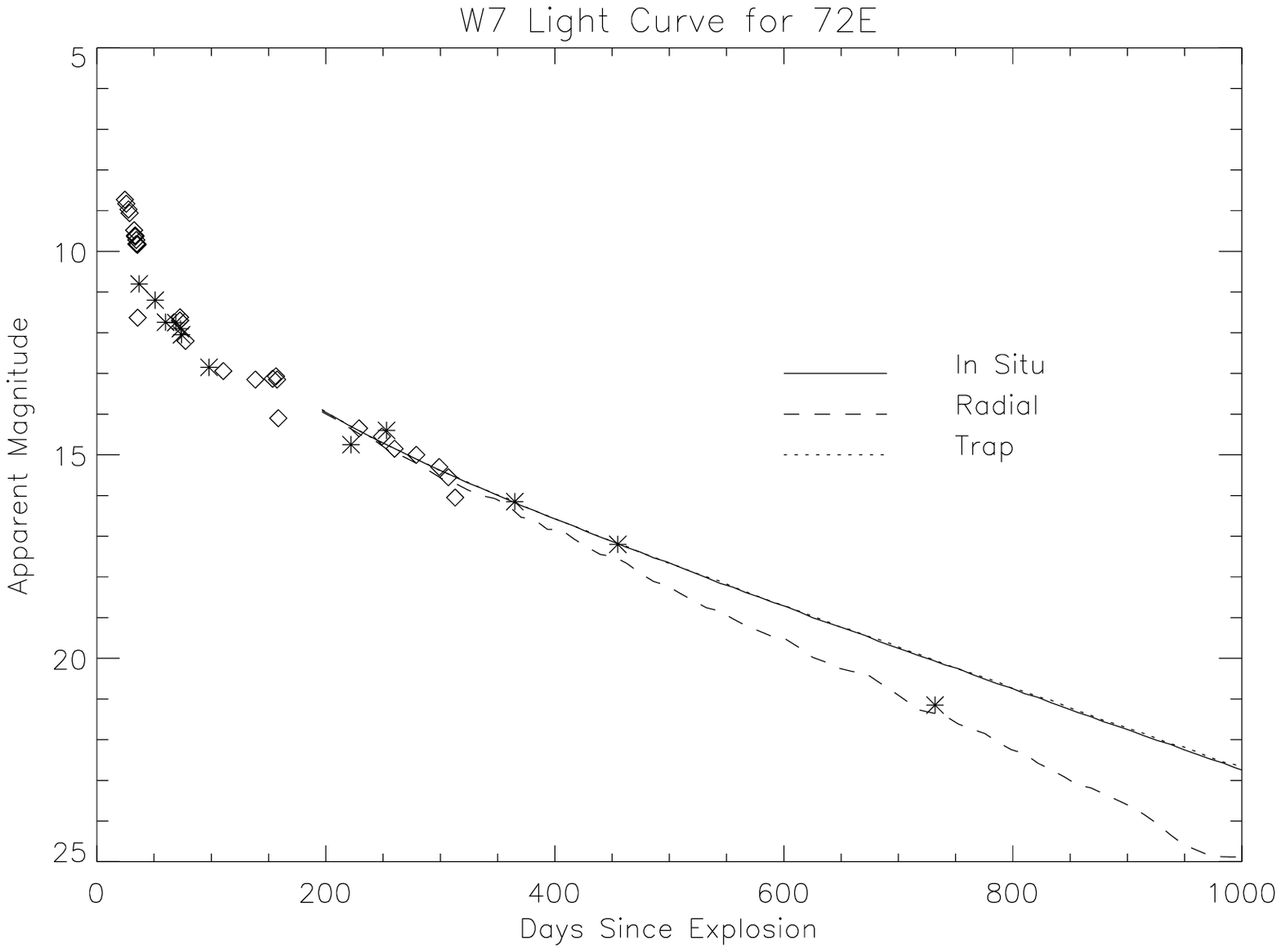,width=2.5in}}
\makebox[2.6in]{\epsfig{file=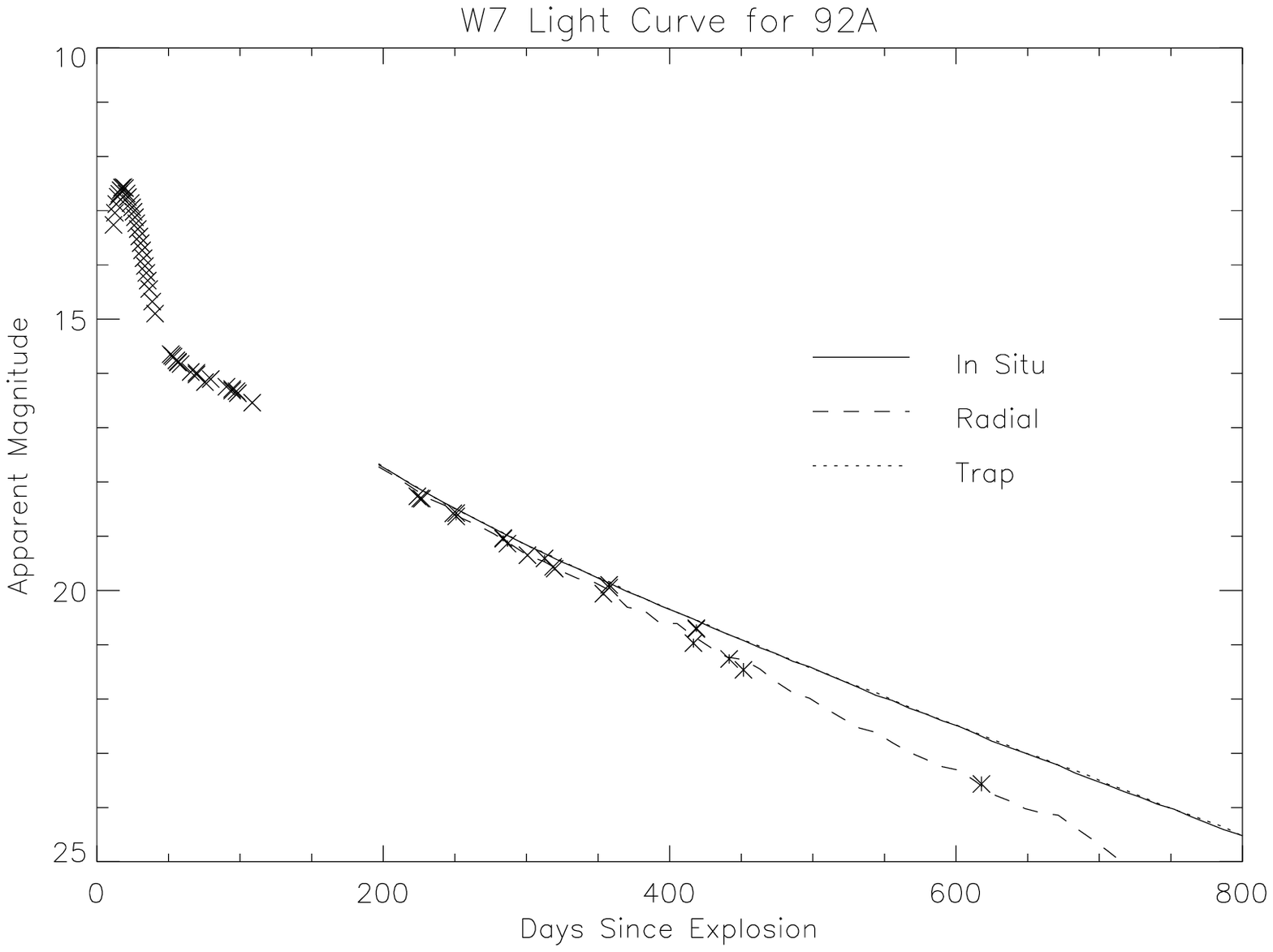,width=2.5in}}

\makebox[2.6in]{\epsfig{file=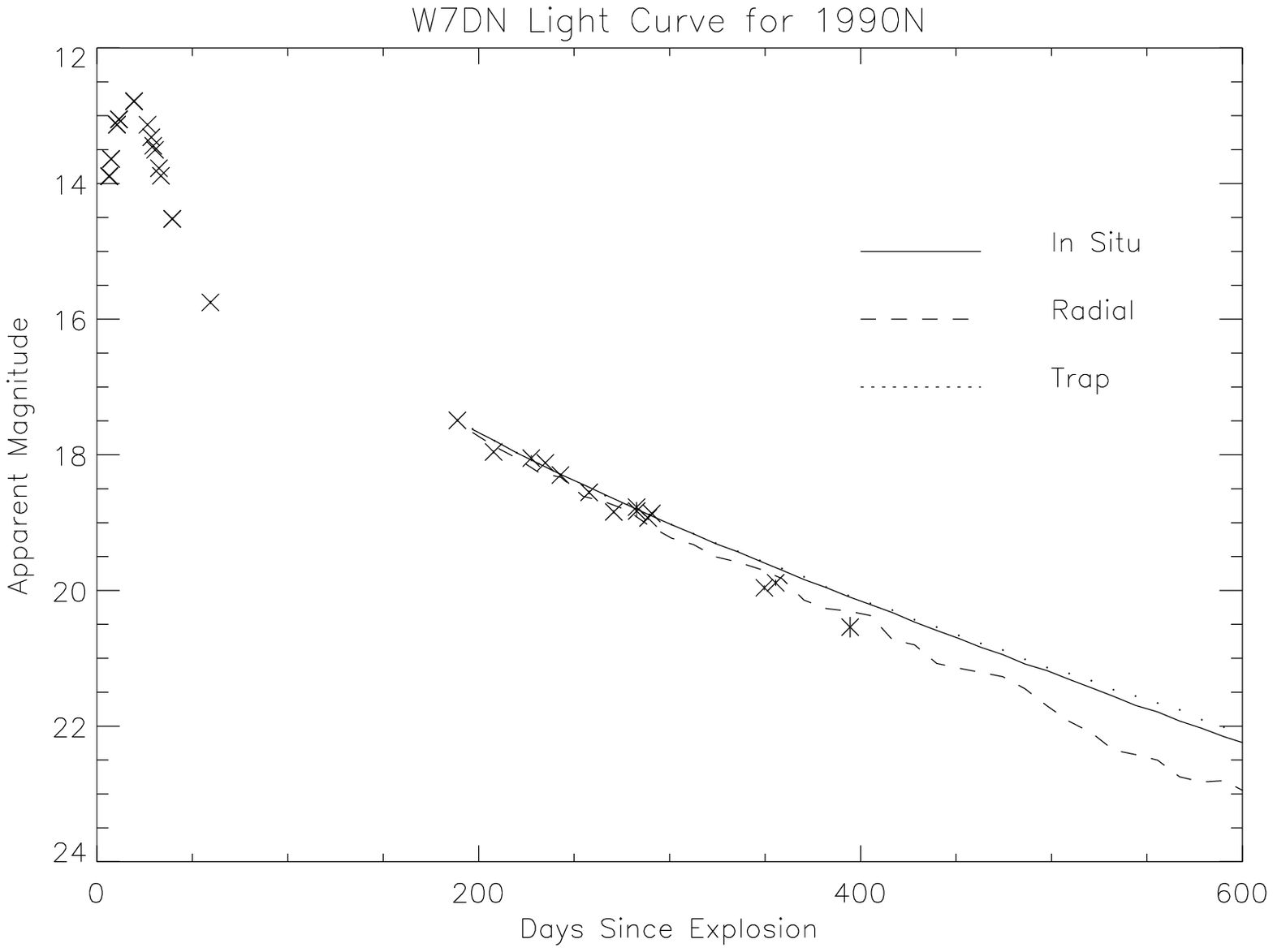,width=2.5in}}
\makebox[2.6in]{\epsfig{file=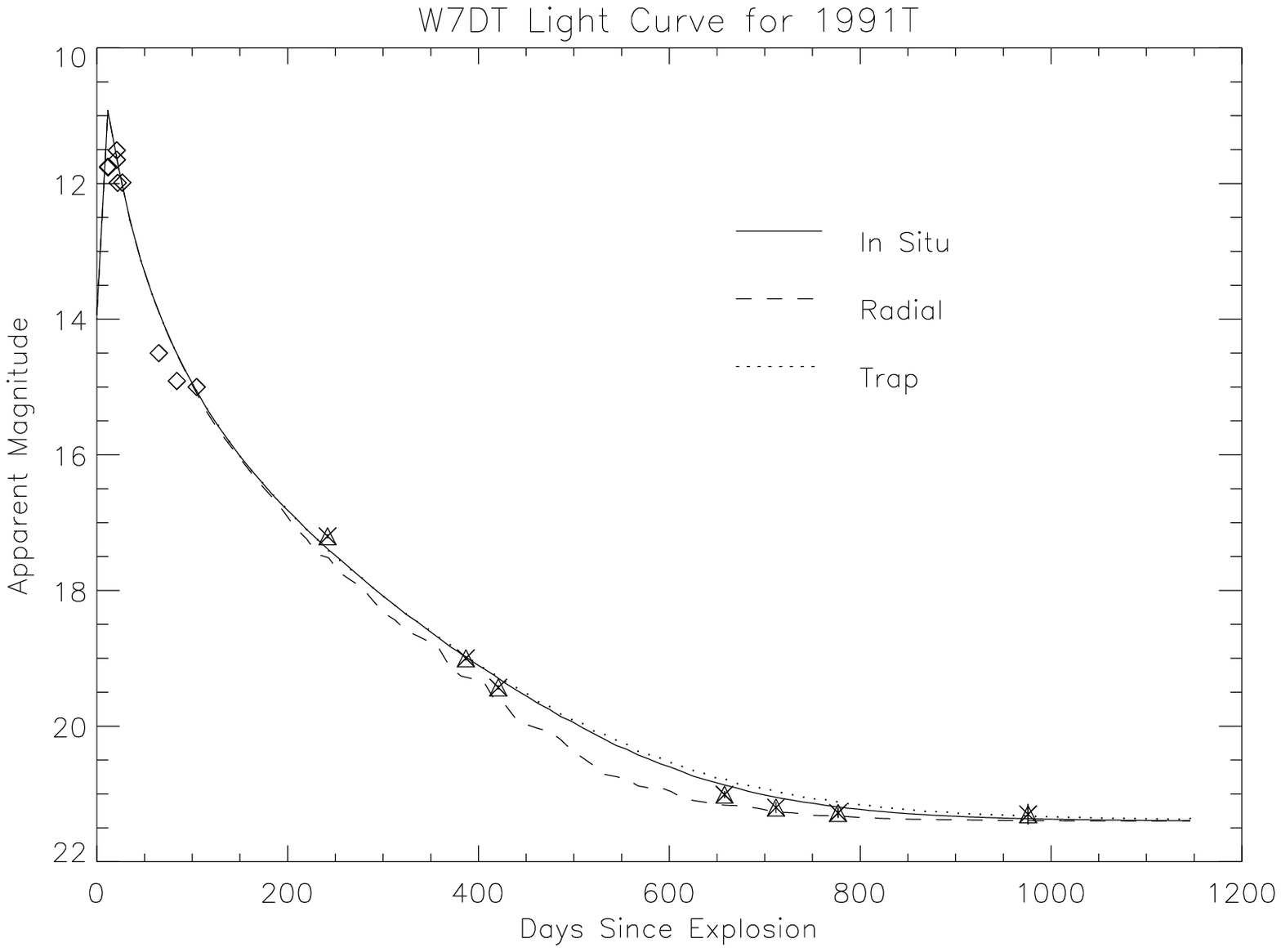,width=2.5in}}
\vspace{10pt}
\caption{Model fits to four SNe observed at late-times. 
See text for references}
\label{fig2}
\end{figure}

SN 1972E, SN 1992A and SN 1990N 
follow linear declines in magnitude after 200 
days, and the slopes of these declines are better fit by the radial models. 
Other models have been suggested as better representatives of these
individual SNe, in future, these other models will be used to strengthen the
argument for these SNe.
 To fit W7DT to
SN
1991T, we included a component with constant flux to account for 
the additional luminosity that is provided
by light echoes, as identified by Schmidt, et.al.(1994)\cite{schmidt94}.
The existence of a light echo provides a luminosity source to offset
the
declining $^{56}$Co input. Although it may still be possible to
accurately model a light curve that contains a light echo, at present
we
caution against their use to discriminate between positron trapping and
escape.

\section*{Survival Fractions and 511 keV Flux}

\begin{table}[h!]
\caption{Positron survival 3 years post-explosion.}
\label{table1}
\begin{tabular}{lllll}
Model Name& Radial \%& Radial \#& Trapped \%& Trapped \# \\
\tableline
W7   &  5.5 & 1.3 x 10$^{53}$ & 2.0 x 10$^{-2}$ & 4.8 x 10$^{50}$ \\
W7DN & 10.4 & 2.6 x 10$^{53}$ & 5.2 x 10$^{-2}$ & 1.3 x 10$^{51}$ \\
W7DT & 11.7 & 3.6 x 10$^{53}$ & 5.9 x 10$^{-2}$ & 1.8 x 10$^{51}$    
\end{tabular}
\end{table}

Table 1 shows the survival fraction of positrons emitted in the SN
models W7, W7DN and W7DT for the two field geometries. The values are
for 3 years (99.99\% of the positrons have been emitted by then). 
Positrons that escape and/or survive until three years after the
supernova explosion exist in an environment that is very diffuse.
This is due to two facts. First, the rapid expansion of the supernova
creates a tenuous bubble, and
second, Type Ia SNe tend to occur in a very low density phase of the
ISM.
This tenuous environment means that a positron can
survive so long
that the contributions from many supernovae can
combine to produce a diffuse positron annihilation radiation source.
Numerically, if the galactic 511 keV flux is 2.3 x 10$^{-3}$
$\gamma$ cm$^{-2}$ s$^{-1}$\cite{share90},
 then a Type Ia SN rate of 0.2 -1 SN Ia [100 y]$^{-1}$ 
would produce the necessary number of positrons,
for
a survival fraction of 5\% from the SN
model W7.

\section*{Summary}

This paper motivates the use of late-time light curves as a probe of
positron transport in type Ia SN ejecta. A few SNe have been observed
well enough to 
begin to discriminate between radial and trapping field geometries.
Preliminary
observations suggest that the field lines are radially
combed, but future observations are required to solve this problem. 

Determining the positron escape from typical Type Ia SNe does not solve
the 511 keV emission problem by itself. The issues of
the galactic distribution of past Ia SNe as well as of the ISM transport and
the
annihilation of positrons must also be addressed. Nonetheless, 
it appears that 
Type Ia
SNe must be considered a potentially large contributor to the
511 keV emission.

\end{document}